\documentstyle[12pt]{article}

 \newcommand{\be}{\begin{equation}}
 \newcommand{\ee}{\end{equation}}
 \newcommand{\ba}{\begin{eqnarray}}
 \newcommand{\ea}{\end{eqnarray}}
 
 \newcommand{\del}{\partial}

\newcommand{\lef}{\left}

\newcommand{\ri}{\right}

\newcommand{\ab}{\bar{A}}

\newcommand{\fb}{\bar{F}}

\newcommand{\ca}{{\cal A}}

\newcommand{\cb}{{\cal B}}

\newcommand{\cf}{{\cal F}}

\newcommand{\cl}{{\cal L}}

\newcommand{\fr}{\frac}

\begin{document}

\begin{titlepage}

\topmargin -15mm

\vskip 10mm
\vskip 25mm

\centerline{ \LARGE\bf Quantum Skyrmions and the Destruction }
\vskip 2mm
\centerline{ \LARGE\bf of Antiferromagnetic Order  }
\vskip 2mm
\centerline{ \LARGE\bf in High-Temperature Superconductors}

    \vskip 2.0cm

    \centerline{\sc E.C.Marino }

     \vskip 0.6cm
     
\centerline{\it Instituto de F\'\i sica}
\centerline{\it Universidade Federal do Rio de Janeiro } 
\centerline{\it Cx.P. 68528, Rio de Janeiro, RJ 21945-970, Brazil} 
\vskip 2.0cm

\begin{abstract} 

A continuous theory which describes the process of doping in high-T$_c$
superconductors of the YBCO-type is presented. It is shown that
the introduction of dopants produces the creation of skyrmion
topological excitations on the antiferromagnetic background. The energy
of these excitations is evaluated. Destruction of the N\'eel ground state
occurs for a critical value $\delta_C$ of the doping parameter for which
the skyrmion energy vanishes. We evaluate this at zero
temperature, obtaining $\delta_C = 0.39 \pm 0.03$ which 
agrees with the experimental value $\delta_C^{exp} = 0.41 \pm 0.02$.

\end{abstract}

\vskip 10mm


\vskip 40mm
Work supported in part by CNPq-Brazilian National Research Council.
     E-Mail address: marino@if.ufrj.br

\end{titlepage}

\hoffset= -10mm

\leftmargin 23mm

\topmargin -8mm
\hsize 153mm
 
\baselineskip 7mm
\setcounter{page}{2}

There is by now clear experimental evidence \cite{ed,ak,ts,rm} that the
{\it pure} high-T$_c$ superconducting cuprates are well described by a
quasi two-dimensional Heisenberg antiferromagnet on a quasi-square lattice,
whose sites are occupied by $Cu^{++}$ magnetic ions.
The continuous version of this \cite{ha}
is the O(3) nonlinear sigma field theory in three-dimensional spacetime
and the ordered magnetization
corresponds to the nonlinear sigma field $n^a, a=1,2,3$. This is subject
to the constraint $n^a n^a = \rho_0^2$, where  $\fr{\hbar c}{a} =
\rho_0^2  \fr{2 \sqrt{2}}{\sqrt{S(S+1)}}$,
$S$ being the spin, $c$ the spin-wave velocity
and $a$, the lattice spacing \cite{ha}. Using linear spin-wave theory
results for the Heisenberg model, on the other hand, we find \cite{ak}
$(\fr{\hbar c}{a})_{LSW} = 1.18 (2S) \sqrt{2} J$
where $J$ is the Heisenberg antiferromagnetic coupling. These two relations
are valid for large $S$. We assume the ratio of both holds for any $S$ and,
in particular, for $S=1/2$ we establish
\be
\rho_0^2 = 1.18 \fr{\sqrt{3}}{4} J
\label{1}
\ee
In this work we start from this continuous description of the pure
system and propose a model for the doping process in YBCO-type materials,
already in the continuous version. A picture emerges, then, that the doping
process produces the formation of skyrmion topological defects on the
antiferromagnetic background, thereby reducing the ordered magnetization.
The skyrmion energy for an arbitrary doping is obtained 
through the evaluation of the quantum skyrmion operator correlation function
and subsequent analysis of its large distance decay. The skyrmion energy
works as an order parameter for antiferromagnetic order and this is
destroyed when that vanishes. This allows us to obtain the critical value
of the doping parameter at zero temperature.

The nonlinear sigma field is
conveniently expressed in the so called $CP^1$ language in terms of a
doublet of complex scalar fields $z_i, i=1,2$ subject to the constraint
$z_i^\dagger z_i = \rho_0^2$, namely
\be
n^a = \frac{1}{\rho_0} z_i^\dagger \sigma^a_{ij} z_j
\label{0}
\ee
where $\sigma^a $ are Pauli matrices. In this language, the continuous
field theory corresponding to the Heisenberg antiferromagnet is
described by the lagrangian density in 2+1 dimensions \cite{ha,raj}
\be
\cl_{CP^1} = (D_\mu z_i)^\dagger (D^\mu z_i) 
\label{2}
\ee
where $D_\mu = \del_\mu + i \ca_\mu$ and $\ca_\mu = \frac{1}{\rho_0^2}
(i z_i^\dagger \del_\mu z_i)$.

The $CP^1$-nonlinear sigma model possesses classical topologically
nontrivial solutions called skyrmions. These bear one unit of the
topological charge $Q= \int d^2x J^0$, where $J^\mu$ is the identically
conserved topological current. In $CP^1$ language $J^\mu =\fr{1}{2\pi}
\epsilon^{\mu\alpha\beta} \del_\alpha \ca_\beta$ and we see that $Q$ is
nothing but the ``magnetic'' flux of the field $\ca_\mu$. This means
that, in this language, the skyrmions are vortices. In the nonlinear
sigma language, they are defects in the ordered N\'eel state.
A general method of vortex quantization \cite{em1,em2}
which allows the explicit construction of a vortex/skyrmion creation operator
$\mu(x)$ therefore, can be applied
to the skyrmions appearing in this theory. Before doing it (Eq. (\ref{12})),
let us consider the process of doping. In order to describe it we already
start with the continuum language and associate to the electron/hole dopants
a two-component Dirac field with two internal indices corresponding to the
spin orientations: $\psi_a (x), a=\uparrow, \downarrow$.
For $YBa_2Cu_3O_{6+\delta}$ the
Fermi surface with dopants has an almost circular shape \cite{ak} and
expanding around it we can use a relativistic approximation for the
dispersion relation. This is clearly not valid for
$La_{2-\delta}Sr_\delta CuO_4$ because of
the shape of the Fermi surface \cite{ak}. We propose
the following lagrangian density for describing the 
dopants and their interaction with the backgroud
lattice in $YBa_2Cu_3O_{6+\delta}$
\be
\cl_{z,\psi} = (D_\mu z_i)^\dagger (D^\mu z_i)
+ i \bar \psi_a \not\! \del \psi_a
-\fr{m^* v_F}{\hbar} \bar \psi_a \psi_a - \bar \psi_a \gamma^\mu \psi_a
\ca_\mu
\label{4}
\ee
where $m^*$ and $v_F$ are respectively the effective mass and Fermi velocity
of dopants. This must be supplemented by the constraint
$z_i^\dagger z_i = \rho_0^2$.
The dopant dispersion relation corresponding to (\ref{4}) is $\epsilon(p) =
\sqrt{p^2 v_F^2 + (m^* v_F^2)^2 }$ which agrees with the fact that the
system is still an insulator even after the inclusion of some doping.
The spin of the dopants is $\vec S = \bar \psi_i \vec \sigma_{ij} \psi_j$.
Notice that the ordered spin of the $Cu^{++}$
electrons is given by (\ref{0}) and
their coupling to the $\ca_\mu$-field is described by
$i z_i^\dagger \stackrel{\leftrightarrow}{ \del}_\mu z_i \ca^\mu$. This
suggests the minimal coupling of the dopants to $\ca_\mu$,
introduced in (\ref{4}).

The doping parameter $\delta$ is introduced by means of a constraint in
the fermion integration measure, namely,
\be
D \lef [ \bar \psi_a,  \psi_a \ri ] =
D \bar \psi_a D \psi_a \delta \lef [ \bar \psi_a \gamma^\mu \psi_a
- \Delta^\mu \ri ]
\label{5}
\ee
where $\Delta^\mu = \delta\ \  \int_{X,L}^\infty d\xi^\mu \delta^3(z-\xi)$
for a dopant introduced at the point $X = (\vec X,T)$ and moving along the
line $L$. The $0^{th}$-component, for instance, is $\Delta^0 (\vec z,t)
= \delta\ \ \delta^2 (\vec z - \vec X(t))$. The above constraint can be
implemented by integration over a vector Lagrange multiplier field
$\lambda_\mu$ coupled as
\be
\cl_\lambda = \lambda_\mu \lef [  \bar \psi_a \gamma^\mu \psi_a
- \Delta^\mu \ri ]
\label{55}
\ee

In order to complete our description of doping, we must introduce their
electromagnetic interaction. The real electromagnetic interaction of
particles of charge $e$ and associated to a current
$j^\mu = \bar \psi_a \gamma^\mu \psi_a $ in 2+1
dimensions is governed by the lagrangian density \cite{em3}
\be
\cl_{EM} = -\fr{1}{4} F_{\mu\nu} \lef [ \fr{1}{\sqrt{-\Box}} \ri ] F^{\mu\nu}
-e \bar \psi_a \gamma^\mu \psi_a A_\mu
\label{6}
\ee
which, therefore, must be included.
   
Before collecting the pieces $\cl_{z,\psi}, \cl_{\lambda}$ and
$\cl_{EM}$, let us remember that
the space of configurations of the $CP^1$-nonlinear sigma model has
inequivalent topological classes
characterized by the Hopf invariant, $S_H$, and
corresponding to the nontrivial mapping $S^3 \rightarrow S^2$. When
performing a functional integration over the $z_i$-fields, therefore,
we must include in principle
a weight $\exp [i \theta S_H ]$ where $\theta$ is a real
parameter to be determined. This implies the addition of a Hopf lagrangian
density $\cl_H$, which in $CP^1$-language is given by
\be
\cl_H = \fr{\theta}{2} \epsilon^{\mu\alpha\beta} \ca_\mu \del_\alpha
\ca_\beta
\label{7}
\ee
It has been shown that $\theta = 0$ for the pure system \cite{tet}, however,
in the presence of doping one has, in principle to add $\cl_H$ as well.
The complete lagrangian density proposed for the description of
$YBa_2Cu_3O_{6+\delta}$ is therefore
\be
\cl = \cl_{z,\psi} + \cl_{\lambda} + \cl_{EM} + \cl_H
\label{8}
\ee
This must be supplemented by the $CP^1$-constraint on the $z_i$-fields.

We remark at this point that the theory described by (\ref{8}) is only
valid up to the threshold where the system is in the
antiferromagnetic phase corresponding to a N\'eel ground state. Above this
point there is no longer an ordered magnetic moment and therefore 
the $CP^1$-nonlinear sigma constraint is no longer justified. We do not
expect, therefore, that the superconducting transition could be explained by
(\ref{8}) without modifications. Observe also that in the region of validity
of (\ref{8}) the presence of an insulating gap is clearly justified.

We now functionally integrate over the fields $z_i$, $\psi_a$, $\lambda_\mu$
and $A_\mu$, in order to obtain an effective lagrangian for $\ca_\mu$. The
$z_i$-integral is easily done in the approximation where
$|z_i|^2 = \rho_{0i}^2 = constant\ \ 
(\rho_{01}^2 + \rho_{02}^2 = \rho_0^2)$ \cite{em2} yielding a piece
$ - \fr {1}{4}\cf_{\mu\nu} [ \fr{\rho_0^2}{-\Box} ] \cf^{\mu\nu} $
for the effective lagrangian density.
The integration over the fermionic fields $\psi_a$ is done by evaluating
the fermionic determinant for an external field
$\bar A = \ca_\mu + \lambda_\mu + e A_\mu$.
In the small effective mass ($m^*$) approximation this yields \cite{cl}
$$
\fr{1}{2} \epsilon^{\mu\alpha\beta} \ab_\mu [\Gamma] \del_\alpha
\ab_\beta -\fr{1}{4} \fb_{\mu\nu} [ \Sigma ] \fb^{\mu\nu}
$$
where $\Gamma$ and $\Sigma$ are kernels whose Fourier transform is
$\Gamma = \fr{1}{2\pi} +\fr{m^* v_F/\hbar}{2\sqrt{k^2}}$ and $\Sigma =
\fr{1}{8\sqrt{k^2}} +\fr{m^* v_F/\hbar}{2\pi (k^2)}$.
The remaining integrations
over $\lambda_\mu$ and $A_\mu$ are quadratic and can be performed exactly
thereby leading to the effective lagrangian density for the $\ca_\mu$-field: 
\be
\cl_\delta [\ca_\mu ] =
\fr{\theta}{2} \epsilon^{\mu\alpha\beta} \ca_\mu \del_\alpha
\ca_\beta - \fr {1}{4} \cf_{\mu\nu} [ \fr{\rho_0^2}{-\Box} ] \cf^{\mu\nu}
- \Delta_\mu \ca^\mu
+ \fr{1}{2} \Delta_\mu \lef [ \Pi P^{\mu\nu} + \Lambda C^{\mu\nu} \ri ]
\Delta_\nu
\label{9}
\ee
where $P^{\mu\nu} = -\Box \delta^{\mu\nu} + \del^\mu  \del^\nu$,
$C^{\mu\nu} = \epsilon^{\mu\alpha\nu}  \del_\alpha$ and $\Pi$ and $\Lambda$
are kernels whose Fourier transform is 
$\Pi = \fr{\alpha + e^2/\hbar c}{(k^2)^{3/2}} - 
\fr{\gamma m^* v_F/\hbar}{k^4}$ and
$\Lambda = \fr{\beta}{k^2} - \fr{\sigma m^* v_F/\hbar}{(k^2)^{3/2}}$.
In the previous expressions, $\alpha, \gamma, \beta, \sigma$ are numerical
coefficients given by $\alpha =\fr{8\pi^2}{\pi^2 + 16}$,
$\gamma =  \fr{32 \pi ( 9\pi^2 -16 )}{ (\pi^2 + 16)^2}$,
$\beta = - \fr{4}{\pi} \alpha$ and 
$\sigma =  \fr{32 \pi^2 ( 24 - \pi^2 )}{ (\pi^2 + 16)^2}$.

The field equation corresponding to (\ref{9}) is
\be
\theta \epsilon^{\mu\alpha\beta}  \del_\alpha \ca_\beta =
\Delta^\mu + \rho_0^2 \lef [ \ca^\mu + \lef ( \fr{1}{-\Box} \ri )
\del^\mu (\del_\nu \ca^\nu )  \ri ]
\label{10}
\ee
Taking the $0^{th}$-component of it and applying to the skyrmion we
see that the last term vanishes and
\be
\theta \epsilon^{ij} \del_i \ca_j = \Delta^0
\ \ 
or
\ \ 
\theta \cb = \delta \ \ \delta^2 (\vec z - \vec X(t))
\label{11}
\ee
where $\vec X(t)$ is the dopant position at a time $t$ and $\cb$ is the
``magnetic'' flux or vorticity of $\ca_\mu$. For the skyrmion $\cb =
\delta^2 (\vec z - \vec X_S(t))$, where $\vec X_S(t)$ is the skyrmion
position. We therefore conclude that
the skyrmion position coincides with the
dopant position at any time and $\theta = \delta$. We conclude
that upon doping, skyrmions are created
on the ordered N\'eel background. This process reduces the ordered magnetic
moment and eventually leads to the complete destruction of the
antiferromagnetic order. A similar situation occurs in
polyacetilene where the doping process produces the formation of
solitons \cite{pol}. Observe that 
$\theta$ vanishes for zero doping, in agreement with \cite{tet}.

Let us evaluate now the quantum skyrmion correlation function. Applying
the formulation developped in \cite{em2,em1} to the system described by
(\ref{9}), we get
\be
<\mu(x) \mu^\dagger(y)>_\delta = Z^{-1} \int D \ca_\mu
\exp \lef \{ \fr{i}{\hbar c} \lef [ S_\delta [\cf_{\mu\nu} +
\tilde B_{\mu\nu} ] + S_{GF} \ri ] \ri \}
\label{12}
\ee
where $S_\delta$ is the action corresponding to (\ref{9}), $S_{GF}$ is
a gauge fixing term and $\tilde B^{\mu\nu} = 2\pi \int_{x,L}^y d\xi_\lambda
\epsilon^{\lambda\mu\nu} \delta^3 (z - \xi)$ is an external field
corresponding to a unit ``magnetic'' flux along $L$ \cite{em2,em1}. In order
to put (\ref{9}) in the form $\cl_\delta [\cf_{\mu\nu} ]$ we wrote,
up to gauge terms $\ca_\mu \rightarrow \fr{- \del_\nu \cf^{\nu\mu}}{-\Box}$.
Also in (\ref{12}), we used $\theta = \delta$ and identified the
skyrmion and dopant positions: $x \equiv X, \ \ y \equiv Y$, according to
(\ref{10}), (\ref{11}) and the discussion thereafter.
Performing the quadratic integration over $\ca_\mu$ in (\ref{12}) and
neglecting the unphysical self-interaction $LL$-terms we get
$$
<\mu(x) \mu^\dagger(y)> = \exp \lef \{ - \fr{1}{\hbar c}
\lef [ \fr{\pi \rho_0^2}{2} - \fr{\delta^2 \gamma m^* v_F c}{8\pi} \ri ]
|x-y| - \fr{\delta^2 (\alpha + e^2/\hbar c)}{4 \pi^2} \ln |x-y| \ri .
$$
\be
\lef .
-i \fr{\delta^2 \beta}{2\pi} [ arg(\vec x - \vec y) + \pi ] \ri \}
\label{13}
\ee
From the large distance behavior of this
we can extract the skyrmion energy for an arbitrary doping:
\be
E_S (\delta) = \fr{\pi \rho_0^2}{2} - \delta^2 \fr{ \gamma \hbar c}{8 a_D}
\label{14}
\ee
In the last expression, we have written the Fermi momentum of the dopants
as \linebreak $m^* v_F = \fr{\hbar \pi}{a_D}$, where $a_D$ is the lattice spacing for
the dopants. The skyrmion energy is an order parameter for the
antiferromagnetic order and we conclude that this is destroyed for a
critical doping parameter
\be
\delta_C = \sqrt{\fr{4 \pi \rho_0^2 a_D}{\gamma \hbar c}}
\label{15}
\ee
The experimental values of the Heisenberg
coupling and spin-wave velocity
for \linebreak $YBa_2Cu_3O_{6+\delta}$ are, respectively,
$J = 100 \pm 20 \ meV $ \cite{sh,ak},
and $c = 1.0 \pm 0.05 \ \fr{eV \AA}{\hbar} $ \cite{rm,ak}.
From $J$, we get $\rho_0^2$  through (\ref{1}) . The only
remaining input is the dopant lattice spacing. We take this to be the
spacing between the oxygen ions in the $CuO_2$-planes,
which for this compound, is $a_D = 2,68$ \AA.
Entering these data and the numerical value
of $\gamma$ (after Eq. (\ref{9})) in (\ref{15}) we get
\be
\delta_C = 0.39 \pm 0.03
\label{16}
\ee
This agrees with the experimental value of the
critical doping at zero temperature \cite{rm}, namely
\be
\delta_C^{exp} = 0.41 \pm 0.02
\label{17}
\ee

We conclude that the continuous theory given by (\ref{9}) provides a good
description of $YBa_2Cu_3O_{6+\delta}$ 
for $\delta \leq \delta_C$, i.e., within the antiferromagnetic ordered
region.
The results also point towards a scenario in which
holes are introduced in the p-orbitals
of the oxygen ions in the $CuO_2$ planes through doping.
For a compound like $La_{2-\delta}Sr_\delta CuO_4$,
for instance, the relativistic
approximation certainly is not valid and we do not expect a quantitative
agreement with the experiment. Nevertheless, it is plausible that the
mechanism of skyrmion formation by doping holds in general and is responsible
for the destruction of the antiferromagnetic order in superconducting
cuprates.
The theory is presumably useful for $\delta > \delta_C$
as well, provided some suitable modifications are introduced. We are
presently investigating this possibility.


\begin{thebibliography}{99}


\bibitem{ed} E.Dagotto, {\it Rev. Mod. Phys.} {\bf 66}, 763 (1994)

\bibitem{ak} A.P.Kampf, {\it Phys. Rep.} {\bf 249}, 219 (1994)

\bibitem{ts} J.M.Tranquada and G.Shirane, in {\it Dynamics of Magnetic
Fluctuations in High-Temperature Superconductors}, G.Reiter, P.Horsch
and G.C.Psaltakis, eds., Plenum, New York (1991)

\bibitem{rm} J.Rossat-Mignod, L.P.Regnauld, J.M.Jurguens, P.Burlet,
J.Y.Henry, G.Lapertot and C.Vettier, in {\it Dynamics of Magnetic
Fluctuations in High-Temperature Superconductors}, G.Reiter, P.Horsch
and G.C.Psaltakis, eds., Plenum, New York (1991)

\bibitem{ha} F.D.M.Haldane, {\it Phys. Rev. Lett.} {\bf 50}, 1153 (1983);
{\it Phys. Lett.} {\bf A93}, 464 (1983)

\bibitem{raj} R.Rajaraman, {\it Solitons and Instantons}, North Holland,
Amsterdam (1982)

\bibitem{em1} E.C.Marino, {\it Phys. Rev. } {\bf D38} 3194 (1988);
{\it Dual Quantization of Solitons} in {\it Applications of Statistical
and Field Theory Methods in Condensed Matter}, D.Baeriswyl, A.Bishop and
J.Carmelo, eds., Plenum, New York (1990)

\bibitem{em2} E.C.Marino, {\it Int. J. Mod. Phys.} {\bf A10}, 4311 (1995)

\bibitem{sh} S.Shamoto, M.Sato, J.M.Tranquada, B.J.Sternlieb and
G.Shirane, {\it Phys. Rev. } {\bf B48}, 13817 (1993)

\bibitem{tet} T.Dombre and N.Read, {\it Phys. Rev.} {\bf B38}, 7181 (1988);
E.Fradkin and M.Stone, {\it Phys. Rev.} {\bf B38}, 7215 (1988);
X.G.Wen and A.Zee, {\it Phys. Rev. Lett.} {\bf 61}, 1025 (1988);
F.D.M.Haldane, {\it Phys. Rev. Lett.} {\bf 61}, 1029 (1988);
L.B.Ioffe and A.I.Larkin, {\it Int. J. Mod. Phys.} {\bf B2}, 203 (1988)

\bibitem{em3} E.C.Marino, {\it Nucl. Phys.} {\bf B408} [FS], 551 (1993)

\bibitem{cl} A.Coste and M.L\"uscher, {\it Nucl. Phys.} {\bf B323}, 631 (1989)

\bibitem{pol} 
H.Takayama, Y.R.Lin-Liu and K.Maki, {\it Phys. Rev.} {\bf B21}, 2388 (1980);
W.Su, J.R.Schrieffer and A.J.Heeger,
{\it Phys. Rev.} {\bf B22}, 2099 (1980);
J.Goldberg {\it et al.},
{\it J. Chem. Phys.} {\bf 70}, 1132 (1979);
S.Ikehata, J.Kaufer, T.Woerner, A.Pron, M.A.Druy, A.Sivak, A.J.Heeger
and A.G.MacDiarmid, {\it Phys. Rev. Lett.} {\bf 45}, 1123 (1980)





\end{thebibliography}
\end{document}